\documentclass[letterpaper,11pt]{article}
\pdfoutput=1

\usepackage{jheppub}
\usepackage{booktabs} 
\usepackage{graphicx} 
\usepackage{siunitx}
\usepackage{cleveref}
\usepackage{epsfig}
\usepackage[normalem]{ulem}
\usepackage{bm}
\usepackage{bbm}
\usepackage{slashed}
\usepackage{xspace}
\usepackage{subfigure}
\usepackage{multirow}

\usepackage{amsmath,latexsym}
\usepackage{pstricks}
\usepackage{color} 
 
\usepackage{marginnote}

\newcommand{\e}{\epsilon}

\newcommand{\nn}{\nonumber}

\newcommand{\beq}{\begin{equation}}
\newcommand{\eeq}{\end{equation}}
\newcommand{\bea}{\begin{eqnarray}}
\newcommand{\eea}{\end{eqnarray}}

\begin{document}


\title{Non-conservative effects on  Spinning Black Holes from World-Line Effective Field Theory}

\author[a]{Walter D. Goldberger,}
\author[a]{Jingping Li,}
\author[b]{Ira Z. Rothstein}
\affiliation[a]{Department of Physics, Yale University, New Haven, CT 06511}
\affiliation[b]{Department of Physics, Carnegie Mellon University, Pittsburgh, PA 15213}

\vspace{0.3cm}


\abstract{

We generalize  the  worldline EFT formalism  developed  in \cite{GnR1,presents,GnR2,Houses,GnR3,GnR4} to calculate the  non-conservative tidal effects on spinning black holes in a long wavelength approximation that is valid to all orders in the magnitude of the spin.   We present results for the rate of change of mass and angular momentum in a background field and find agreement with previous calculations obtained by different techniques. We also present new results for both the non-conservative equations of motion and power loss/gain for a binary inspiral, which start at 5PN and 2.5PN order respectively  and manifest the Penrose process.

}
\maketitle



\section{Introduction}

Understanding the dynamics of spinning black holes (BHs) is of both formal and phenomenological interest~\cite{Thorne:1984mz}.   Although much is known about the quasinormal mode spectrum of the Kerr solution, understanding how those modes react to external perturbations presents a theoretical challenge. External tidal
fields will distort the black hole both by inducing changes in multipole moments as well as changing the mass and angular momentum. Phenomenologically these tidal effects can become sufficiently large to be relevant to parameter estimation in binary inspiral gravitational waveforms. While for non-spinning black holes these finite size are fourth order corrections in the Post-Newtonian (PN) expansion~\cite{Poisson:1994yf,Tagoshi:1997jy}, they are enhanced to 2.5PN order for the case of maximally rotating BHs~\cite{Tagoshi:1997jy}. 
 
In this paper we will utilize worldline EFT techniques~\cite{GnR1,presents,GnR2,Houses,GnR3,GnR4} to calculate non-conservative finite size effects for Kerr BHs, in a limit in which their size is parametrically smaller than the curvature length scale of any external gravitational fields, for arbitrarily large spins.    In order to account for the physics of the horizon within a point particle description, we use a method introduced in ref.~\cite{GnR2} for the case of non-rotating black holes, which attributes dissipation across the horizon to a set of worldline localized modes.    The dynamics of these modes is encoded in a set of correlation functions that can be obtained by a matching calculation to graviton absorption and (in the quantum mechanical case~\cite{Houses,GnR3}) emission processes in the single BH sector.   

Within the context of spinning BHs, the EFT formalism for dissipation was previously discussed first in ref.~\cite{Porto} who considered the tidal couplings of gravitons in the limit of slow rotation, and then in ref.~\cite{Endlich} which also restricted its analysis to slow spin but discussed the couplings to more general external fields.   Here, we generalize these methods to the case of Kerr BHs with arbitrary rotation parameter, including the phenomenologically relevant limit of maximal (near extremal) spin.

We begin in sec.~\ref{sec:lemech} by constructing the Schwinger-Keldysh~\cite{CTP}, or ``in-in" action of a generic spinning compact object tidally coupled\footnote{We restrict ourselves to leading order quadrupolar gravitational interactions in this paper.} to a background gravitational field.   The compact object is described by a worldline $x^\mu$ and a local rotation frame~\cite{Hansons,Porto:2005ac} $e^a{}_\mu$, plus a generic set of internal modes $X$ responsible for dissipation.    The in-in action is obtained by integrating out the internal modes, and its variation yields \emph{instantaneous} (i.e. valid locally at any parameter time along the worldline) equations of motion for the center of mass momentum and spin of the BH.   In the linear response approximation, these equations depend on the retarded Green's function of the internal modes, or equivalently, the induced quadrupole moments (in the sense defined in sec.~\ref{sec:lemech}) due to the external field.  If there is a separation of scales between the internal response and external field time scales, the equations of motion become local.   The resulting equations of motion are quadratic in the background curvature and contain time-reversal even terms corresponding to conservative tidal effects (e.g. static Love numbers), and time-reversal odd parts which encode the effects of dissipation (e.g. tidal heating and torquing).

The results presented in sec.~\ref{sec:lemech} are generic.   In sec.~\ref{sec:match} we specify to the case of Kerr BHs.    The Wightman functions of the EFT are extracted in sec.~\ref{sec:wight} by matching to the low-energy graviton absorption probabilities computed in~\cite{Starobinsky,Page}.   In sec.~\ref{sec:ret}, we determine the retarded Green's function relevant to classical dynamics from these Wightman functions.   To do so, we employ dispersion relation methods, which fix the absorptive (or non-``local", in the sense defined in sec.~\ref{sec:ret}) part, in combination with recent results of~\cite{Chia:2020yla}, indicating that Kerr BHs have vanishing conservative tidal response at low frequency, to fix the local part.   As a check, we compute in sec.~\ref{sec:poi} the rate of change of mass and spin of a Kerr BH placed in long wavelength tidal background field whose average over worldline time are in agreement with previous results in~\cite{death,Poisson:2004cw,poisson2,poisson3} .    

Finally, as a novel application of our methods, we obtain the non-conservative part of the PN equations of motion for spinning black hole binaries in sec.~\ref{sec:PN}.    These include instantaneous forces and torques which, in the near maximal spin case, scale as 5PN and 4PN respectively.   We also compute the rate of change of orbital energy and angular momentum due to horizon finite size effects.    The former effect, arising at 2.5PN order for large spins, is not small compared to finite size effects in neutron stars (5PN) or slowly rotating BHs's (4PN), and may have implications for gravitational waveform modeling.   In sec.~\ref{sec:conc}, we conclude and outline directions for future work.

\section{Dissipative mechanics of spinning compact objects}
\label{sec:lemech}

In this section, we consider how dissipative processes affect the motion of a spinning compact object moving through a fixed background spacetime.   We work in the limit $R\ll {\cal R}$ where object's radius is $R$  and  ${\cal R}$ is the typical length scale over which the background metric varies (the curvature radius).   In this limit, the object may be described by worldline effective field theory \cite{GnR1,presents}, with finite size effects encapsulated by local, curvature dependent terms in a generalized point-particle Lagrangian.

In order to account for dissipative effects while retaining a point particle description, we employ the framework introduced in~\cite{GnR1,presents} and further developed in~\cite{GnR2,GnR3,Houses,GnR4}.   In this approach,  long wavelength dissipative effects due to the internal structure (finite size effects) is attributed to the existence of gapless modes localized on the worldline which absorb energy as well as linear and angular momentum from the external environment.

We begin with the worldline theory in the absence of dissipation.    In addition to the trajectory $x^\mu(\lambda)$, we also introduce an orthonormal frame $e^a{}_\mu$ which is necessary to describe spin dynamics~\cite{Hansons}, 
and obeys the constraints
\beq
\label{eq:econ}
\eta_{ab} e_{\mu}^{a} e_{\nu}^{b}=g_{\mu \nu}(x),  \quad g_{\mu \nu}(x) e_{a}^{\mu} e_{b}^{\nu}=\eta_{ab},
\eeq
where $a,b=0,1,2,3$ are local Lorentz indices.   The rotation of the particle relative to fixed inertial frames is then encoded in the angular velocity 
\beq
\Omega^{ab}= g^{\mu\nu} e^a{}_\mu {D\over D\lambda}  e^b_\nu = - \Omega^{ba},
\eeq
with ${D\over D\lambda} e^a{}_\mu ={\dot x^\rho} \nabla_\rho e^a_\mu= {\dot e}^a{}_\mu - \Gamma^\rho{}_{\sigma\mu} {\dot x}^\sigma e^a{}_\rho$ (the overdot correspond to differentiation with respect to $\lambda$).   Finally, it is convenient to define an einbein $e(\lambda)$ in order to enforce reparameterization invariance $\lambda\mapsto \lambda'(\lambda)$, $e'(\lambda') d\lambda' = e(\lambda) d\lambda$.

We begin by writing down the most general reparameterization invariant (RPI) action to leading order in a derivative expansion in powers of ${ R}/{\cal R}\ll 1$,
\beq
\label{eq:Spp}
S_{pp} = -\int dx^\mu p_a  e^a{}_\mu+ {1\over 2}\int d\lambda~ S^{ab} \Omega_{ab} + {1\over 2} \int ds  \left(p_a p^a -m^2\right) + \int ds \lambda_a S^{ab} p_b+\cdots,
\eeq 
where $ds =e(\lambda) d\lambda$, with units of time/energy, is RPI.  For our purposes here, this treatment is more convenient than the Routhian approach employed in~\cite{PR}.  In this equation the momentum $p_\mu = -\delta S_{pp}/\delta {\dot x}^\mu$ and spin $S^{ab}=\partial S_{pp}/\partial \Omega_{ab}$ are conjugate variables to $x^\mu$ and $e^a{}_\mu$.   The quantity $m^2$ is an arbitrary function of all possible scalars constructed out of $p_a$, $S^{ab}$ and $g_{\mu\nu}$.    The form of this function is not predicted by the point-particle EFT, but rather must be fixed through a matching procedure to the UV theory of the extended object.   $m^2(p,S)$ determines the relation between the spin and angular velocity $\Omega^{ab}$.  It also fixes the Regge trajectory~\cite{Hansons} of the spinning particle, i.e. the relation between the invariant mass $p^2$ and the spin, which follows from variation of $S_{pp}$ with respect to $e(\lambda)$.

The last term in Eq.~(\ref{eq:Spp}), involving the Lagrange multiplier  $\lambda_a$ enforces a `supplementary' constraint on $S^{ab}$ that reduces the number of spin degrees of freedom down to the three required by Poincare symmetry.  We find it convenient to impose
\beq
S^{ab} p_b =0.
\eeq
which is known as the covariant spin supplementary condition, other choices~\cite{pryce,Henneaux} have no effect on physical predictions.   Likewise, $\lambda_a$ itself is ambiguous since it can be shifted by an amount proportional to $p_a$ without affecting the equations of motion.   The variation of $S_{pp}$ with respect to the kinematic variables $(x^\mu,p_a,e^a{}_\mu,S^{ab},\lambda_a,e)$  leads to the Papapetrou-Mathison-Dixon~\cite{spin} equations of motion  for $p_\mu = e^a{}_\mu p_a$ and $S_{\mu\nu} = e^a{}_\mu e^b{}_\nu S_{ab}$, with $p_\mu p^\mu=m^2$.

To account for dissipation, we now include a set of internal worldline degrees of freedom, which we generically label as $X(\lambda)$ in what follows.   The dynamics of the composite object then follow from an action
\beq
\label{eq:SX}
S =  -\int dx^\mu e^a{}_\mu p_a(X) + {1\over 2}\int d\lambda S^{ab}(X) \Omega_{ab} +\int ds(p_a p^a + L_X(X,e^{-1} {\dot X})+ S_{int}+\cdots
\eeq
The internal dynamics is encoded in the Lagrangian $L_X$ whose detailed form will not be needed in order to obtain our results.
In the case of a composite object, the momentum $p_a(X)$ and spin $S^{ab}(X)$ are interpreted as `composite operators' (c.f. a hydrogen atom or relativistic string) which account for the possibility that excitations or de-excitations of the internal modes $X$ can contribute to changes in the linear and angular momenta measured by asymptotic observers.   
Note that this implies that we are not treating $p$ and $S$ as independent degrees of freedom and,  as such, we 
will not vary them in the action. In the non-dissipative case the resulting equations of motion would lead
to a relation between $p/S$ and $\dot x/\Omega$.   Instead,  we will determine the relation between the velocity $dx^\mu/ds$ and the momentum by imposing that the operator constraint $S^{ab} p_b =0$ be a constant of the motion.

The term 
\beq
\label{eq:sint}
S_{int}  =- \int ds Q^E_{ab}(X,e) E^{ab}(x,p)  - \int ds Q^B_{ab}(X,e) B^{ab}(x,p) + \cdots.
\eeq
in Eq.~(\ref{eq:SX}) describes the interaction of the internal modes with an external gravitational field, to leading order in gradients of the spacetime metric.     $Q^E_{ab}$ and $Q^B_{ab}$ are dynamical moments, which are also composite operators built out of the variables $X(\lambda),e(\lambda)$ in some unspecified way, that  couple to the electric and magnetic components of  the of the Weyl  tensor, $W_{abcd}$,  in the local Lorentz frame of the rotating particle,
\bea
E_{ab} &=& W_{acbd} {p^c p^d\over p^2}, \\
B_{ab} &=& {\tilde W}_{acbd} {p^c p^d\over p^2} = {1\over 2} \epsilon_{acd} W^{cd}{}_{eb} {p^e\over \sqrt{p^2}}, 
\eea
where we have defined $\epsilon_{abc} = \epsilon_{dabc} p^d/\sqrt{p^2}$ as the Levi-Civita tensor transverse to $p^a$.
Note that we have not included any explicit couplings  to the Ricci curvature, as such terms can be removed by field redefinitions of the spacetime metric and therefore have no physical content.   For the same reason, only the traceless, transverse to $p^a$, components of the tensors $Q^{E,B}_{ab}$ couple to the external field.  

In order to determine how the internal modes $X$ affect the time evolution of the kinematic variables, we employ the ``in-in'' (or Schwinger-Keldysh) closed time path~\cite{CTP} of quantum mechanics.   This time asymmetric  approach allows one to treat dissipative systems within an action formalism.
  In the case of black holes, the quantum mechanical nature of the underlying degrees of freedom is not important for the classical processes described here.   Nevertheless, as discussed in~\cite{GnR4}, a quantum formulation of the black hole EFT (including such effects as Hawking radiation~\cite{GnR3}) can be used to efficiently describe black hole dissipation even in the purely classical regime.

The Schwinger-Keldysh effective action ($\Gamma$) is defined through
\beq
\exp\left[{i\Gamma[\Delta,e;{\tilde\Delta},e]}\right] = \int D X D {\tilde X} \exp\left[iS[\Delta,X,e]-iS[{\tilde \Delta},{\tilde X},{\tilde e}],\right],
\eeq
 Here, the path integral is over two copies of the $X$ degrees of freedom with fixed boundary conditions at initial time.   Integration over $X,{\tilde X}$ yields a functional $\Gamma[\Delta,e;{\tilde \Delta},{\tilde e}]$ whose variation determines the classical motion of $\Delta=(x^\mu,e^a{}_\mu)$ and the einbein $e$:
\beq
\label{eq:dG}
\left.{\delta\over \delta \Delta(\lambda)}\Gamma[\Delta,e;\tilde{\Delta},{\tilde e}]\right|_{\Delta={\tilde\Delta};e={\tilde e}} =0.
\eeq
Note that by construction, $\Gamma[\Delta,e;{\tilde\Delta}=\Delta,{\tilde e}=e]=0$.   Treating the interaction terms with the background curvature in Eq.~(\ref{eq:SX}) perturbatively, the variation of  $\Gamma[\Delta,e;{\tilde \Delta},{\tilde e}]$ relates the equations of motion for the orbital degrees of freedom to the correlation functions of the operators $Q^{ab}_{E,B}$, which can be calculated by matching to a full theory description of the internal structure of the compact objects.

Variation of the in-in action $\Gamma$ with respect to the einbein $e(\lambda)$ yields 
\beq
\label{eq:ms}
\langle p_a p^a -H_X(X) - H_{int}\rangle = 0.
\eeq
%
%
In this equation, $\langle\cdots\rangle$ denotes a quantum expectation value in the initial state of the internal modes $X(\lambda)$, and corresponds to the in-in path integral expression
\beq
\label{eq:VEV}
\langle {\cal O}[X]\rangle = \int D X D {\tilde X} e^{iS[\chi,X,e]-iS[{\tilde \chi},{\tilde X},{\tilde e}]} {\cal O}[X]
\eeq
for any composite operator ${\cal O}$.    The internal Hamiltonian, in the absence of interactions,  is 
\beq
H_X = -{\delta\over\delta e} \int ds L_X(X,e^{-1} {\dot X}) = {\dot X} {\partial L_X\over \partial {\dot X}} - L_X,
\eeq
while the tidal coupling gives 
\beq
H_{int} = -{\delta\over\delta e}\int d\lambda e \left(Q^E_{ab} E^{ab}+Q^B_{ab} B^{ab}\right).
\eeq
In the absence of external curvature, $H_{int}=0$ and Eq.~(\ref{eq:ms}) has the interpretation of a mass-shell constraint $\langle H_X(X)\rangle =\langle p_a p^a\rangle$ whose solution relates the invariant mass $M^2=\langle p_a p^a\rangle$ to the initial state of the variables $X$.   More generally, when $H_{int}\neq 0$, Eq.~(\ref{eq:ms}) determines in principle how $M^2$ changes as a result of tidal interactions with the external field.


The change in $M^2$ due to tidal interactions can also be obtained from the equation of motion for $ p^\mu =e^\mu{}_a \langle p^a(X)\rangle,$ which follows from the variation of the Schwinger-Keldysh action with respect to $x^\mu(\lambda)$.  To perform this variation, it is convenient to work in Fermi normal coordinates centered on the worldline, i.e. $\partial_\sigma g_{\mu\nu}(x(\lambda))  = 0$, and then covariantize to obtain a result valid in any coordinate system.    This yields,
\bea
\label{eq:dpdt}
{D\over D s} p^\mu = -{1\over 2} R^\mu{}_{\lambda\rho\sigma} {dx^\lambda\over ds} S^{\rho\sigma} + e^a{}_\rho e^b{}_\sigma\left[\langle Q^E_{ab}\rangle \nabla^\mu E^{\rho\sigma} + \langle Q^B_{ab}\rangle \nabla^\mu B^{\rho\sigma}\right], 
 \eea    
  where we have defined (by an abuse of notation), $\nabla_\mu E_{\rho\sigma} = p^\alpha p^\beta (\nabla_\mu W_{\rho\alpha\sigma\beta})/p^2,$ $\nabla_\mu B_{\rho\sigma} = p^\alpha p^\beta (\nabla_\mu {\tilde W}_{\rho\alpha\sigma\beta})/p^2$.    The first term\footnote{Note that because of the constraints Eq.~(\ref{eq:econ}) on $e^a{}_\mu$, the variation of the action with respect to $x^\mu$ must be compensated by the variation $\delta e^a{}_\mu = {1\over 2} g^{\rho\sigma} \delta g_{\sigma\mu}, \delta g_{\mu\nu} = \delta x^\rho\partial_\rho g_{\mu\nu}$ in order to generate the first term in Eq.~(\ref{eq:dpdt}).} on the RHS corresponds to the usual Mathison-Papapetrou-Dixon~\cite{spin} force on a spinning point particle, where $S^{\mu\nu}= \e^\mu{}_a e^\nu{}_b \langle S^{ab}(X)\rangle$ is the physical spin of the object as measured by asymptotic observers, while the remaining terms give the finite size corrections.  
  
The expectation values $\langle Q^{E,B}_{ab}\rangle$ in Eq.~(\ref{eq:dpdt}) are defined through Eq.~(\ref{eq:VEV}) and are in general functionals of the applied fields $E_{ab},B_{ab}$ as well as the orbital degrees of freedom.    For weak external fields, linear response theory implies that the the expectation values $\langle Q^{E,B}_{ab}\rangle$ are of the form\footnote{Parity invariance forbids a term in linear in $B_{ab}$ from appearing in $\langle Q_{ab}^E\rangle$ and similarly for $\langle Q_{ab}^B\rangle$}
\beq
\langle Q_{E}^{ab}(s)\rangle = \int ds' G^{ab,cd}_{R,E}(s-s') E_{cd}(x(s'))+{\cal O}(E^2),
\eeq
and similarly for $\langle Q^B_{ab}\rangle$, where the retarded Green's function is
\beq
G^{ab,cd}_{R;E,B}(s-s') = - i \theta(s-s') \langle [Q_{E,B}^{ab}(s),Q_{E,B}^{cd}(s')]\rangle,
\eeq
with the expectation value calculated at zero external field, in the initial state of the compact object\footnote{For a classical black hole, this state is labeled by mass, spin and electric plus magnetic charges.}.  If the internal dynamics is fast compared to the time scale of the tidal perturbation, the response can be regarded to be nearly instantaneous, ie. a sum of time derivatives of delta functions, with coefficients that depend on the internal structure of the compact object, of the form 
\beq
\langle Q^E_{ab}(s) \rangle\approx \Lambda_0^{ab,cd} E_{cd}(x(s)) + \Lambda_1^{ab,cd}{d\over ds}E_{cd}(x(s))+\cdots.
\eeq   
The tensors $\Lambda_{0,1}^{ab,cd}$, which can depend on $p^a=\langle p^a(X)\rangle$ and  $S^{ab}=\langle S^{ab}(X)\rangle$, carry information about the microscopic structure of the compact object.   In particular, $\Lambda_{0}^{ab,cd}$ encodes the static response (including possible Love numbers) while $\Lambda_{1}^{ab,cd}$ includes the effects of dissipation.   We will compute $\Lambda_{1}^{ab,cd}$ for the Kerr BH in sec.~\ref{sec:match}.  Note that the  derivative of the curvature includes both the intrinsic time dependence of the background field as well as that induced by the rotation of the compact object:
\beq
\label{eq:edot}
{d\over ds} E_{ab} = e^{-1} {d\over d\lambda} E_{ab} = e_a{}^\mu e_b{}^\nu  \left({dx^\rho\over ds} \nabla_\rho\right) { E}_{\mu\nu}-e^{-1}\Omega_a{}^c E_{cb} -e^{-1}\Omega_b{}^c E_{ac}.
\eeq

 To obtain the spin equation of motion, we vary the in-in effective action with respect to the frame $e^a{}_\mu$.   This variation must be carried out in a way that preserves the constraints \cite{Hansons} in Eq.~(\ref{eq:econ}), i.e $\delta e^a{}_\mu = \theta^a{}_b e^b{}_\mu$, with parameters $\theta^{ab} = -\theta^{ba}$ and yields the evolution equation for the spin $S^{\mu\nu} = e^\mu{}_a e^\nu{}_b \langle S^{ab}(X)\rangle$, 
\bea
\label{eq:dsdt}
{D\over Ds} S^{\mu\nu}&=& {dx^\nu\over ds} p^\mu -{dx^\mu\over ds} p^\nu + 2 e^\mu{}_a e^\nu{}_b \left[\langle Q^E_{cd}\rangle {\delta \over \delta \theta_{ab}} E_{cd}+\langle Q^B_{cd}\rangle {\delta \over \delta \theta_{ab}} B_{cd}\right],
\eea
with variations
 \bea
{1\over 2} \langle Q^E_{cd}\rangle {\delta\over\delta\theta^{ab}} E^{cd} &=&  \langle Q^E_{c[a}\rangle E^c{}_{b]}  - \langle Q^E_{cd}\rangle {p_{[a}\epsilon_{b]}{}^{ce} B^d{}_e\over\sqrt{p^2}} ,\\ 
 {1\over 2} \langle Q^B_{cd}\rangle {\delta\over\delta\theta^{ab}} B^{cd}&=& \langle Q^B_{c[a}\rangle B^c{}_{b]}  - \langle Q^B_{cd}\rangle {p_{[a}\epsilon_{b]}{}^{ce} E^d{}_e\over\sqrt{p^2}},
 \eea
that can be deduced from the electric-magnetic (Bel) decomposition of the Weyl tensor~\cite{hawkeb},
\beq
W^{ab}{}_{cd} = -4 E^{[a}{}_{[c} \delta^{b]}{}_{d]} + {8\over p^2} E^{[a}{}_{[c} p^{b]} p_{d]} +{4\over \sqrt{p^2}} \epsilon^{ab}{}_{e} B^e{}_{[c} p_{d]} +  {4\over \sqrt{p^2}} \epsilon^{cd}{}_e  B^{e[a} p^{b]}.
\eeq

To complete the system of equations of motion we need relations between the pairs $(p_\mu,S^{\mu\nu})$ and $(\dot x^\mu, \Omega^{\mu\nu})$, which in the absence of dissipation, are usually determined by varying the action with respect to $p_\mu,S^{\mu\nu}$.     However, in Eq.~(\ref{eq:SX}), the momentum variables are regarded as dependent on the internal modes $X$, so the necessary relations would follow from the equations of motion for $X$.    Because these equations depend on the (unknown) Lagrangian $L_X(X,e^{-1} {\dot X})$, it is not possible to obtain them in a model-independent way.    For the Kerr BH, we will sidestep this issue in the next section by performing an explicit matching calculation that yields the relation between $S^{\mu\nu}$ and $\Omega^{\mu\nu}$.     On the other hand, the relation between $\dot x^\mu$ and the momenta can be obtained without knowledge of the Lagrangian, by demanding that consistency of the  constraint $S^{ab} p_ b=0$ with the equations of motion Eq.~ (\ref{eq:dsdt}), (\ref{eq:dpdt}).   This yields
\bea
{dx^\mu\over ds} +{1\over 2p^2} R_{\nu\lambda\rho\sigma} {dx^\lambda\over ds}  S^{\mu\nu} S^{\rho\sigma}-  \frac{p^\mu}{p^2}  p\cdot \frac{dx}{ ds}   &= &{1\over p^2}\langle Q^E_{ab}\rangle\left(2\sqrt{p^2} e^\mu{}_d \epsilon^{dac} B^b{}_c+ e^a{}_\rho e^b{}_\sigma S^{\mu\nu} \nabla_\nu E^{\rho\sigma}\right) \nn \\
\nn
& &+ {1\over p^2}\langle Q^B_{ab}\rangle\left(2\sqrt{p^2} e^\mu{}_d \epsilon^{dac} E^b{}_c+ e^a{}_\rho e^b{}_\sigma S^{\mu\nu} \nabla_\nu B^{\rho\sigma}\right),\\
\label{eq:pvsx}
\eea
which can be used to relate $p^\mu$ to $dx^\mu/ds$.   In particular, for an object at rest in the absence of background fields, this equation simply states that $p^\mu$ and $dx^\mu/ds$ are colinear.   Therefore, in terms of the proper time $\tau$ along the worldline, the four-velocity obeys the usual relation $v^\mu = dx^\mu/d\tau= p^\mu/M$.

In the case of a point-like spinning particle without tidal interactions, the equations of motion imply that $p^2=M^2$ and $S^2\equiv {1\over 2} S_{\mu\nu} S^{\mu\nu}$ are conserved along the worldline.   The main new feature of the composite object is that it incorporates the effects of dissipation, for instance the accretion/loss of mass and spin due to interactions with the background curvature.   Including dissipative effects, we have instead (up to cubic and higher powers of the background curvature)
\bea
\label{eq:dmdt}
{d\over ds} M^2= 2 e^a{}_\rho e^b_\sigma\left[\langle Q^E_{ab}\rangle (p\cdot\nabla) E^{\rho\sigma} +  \langle Q^B_{ab}\rangle (p\cdot\nabla) B^{\rho\sigma}\right]\neq 0,
\eea
Similarly, the tidal coupling in Eq.~(\ref{eq:dsdt}) can generate non-trivial time dependence for $S^2$: 
\beq
\label{eq:ds2dt}
{d\over ds} S^2 = 4 \langle Q^E_{ab} \rangle E^{bc} S^a{}_c + 4 \langle Q^B_{ab} \rangle B^{bc} S^a{}_c\neq 0.
\eeq 
We will use these results in sec.~\ref{sec:poi}, sec.~\ref{sec:PN} to obtain predictions for the rate of mass and spin dissipation of black holes coupled to tidal backgrounds.

\section{EFT matching for spinning black holes}
\label{sec:match}

In this section we apply the general framework described in the previous section to the case of a Kerr black hole with mass $M$ and spin $S$.    For a BH probed by fields with characteristic frequency $\omega$, the EFT power counting consists of a double expansion in the small parameters $\kappa\equiv \hbar \omega/M_{Pl}\ll 1$ and $G_N M \omega \ll 1$.    The parameter $\kappa$ controls quantum gravity effects, which are negligibly small for the applications considered in this paper, while $G_N M \omega \ll 1$ is an expansion parameter for  finite size effects.   Our EFT can be used to describe black holes with arbitrary values of the dimensionless rotation parameter $\chi= S/G_N M^2$ in the full range $\chi^2\leq 1,$ including the near extremal (maximally rotating) case with $\chi\sim {\cal O}(1)$.   Equivalently, our EFT is valid regardless of the size of the ratio $\omega/\Omega_H$ (with $\Omega_H$ the angular velocity of the horizon), and therefore extends previous work~\cite{Porto,Endlich} on worldline EFTs for spinning BHs, allowing us to make predictions for non-conservative effects in the regime $\Omega_H\gg\omega$ where these effects are enhanced relative to the non-spinning case.   We will only work to  leading order in the power counting, although the formalism allows for systematic corrections.

\subsection{Wightman functions}
\label{sec:wight}

Here we extract the two-point Wightman correlators of the composite operators $Q^{E,B}_{ab}$, by matching to the  graviton absorption probability given in~\cite{Starobinsky,Page}.   The incoming graviton is taken to be in a state with fixed angular momentum quantum numbers $(\ell,m,h=\pm 2)$, sharply localized about a frequency $\omega$.   Classically, the probability is simply the coefficient for absorption of an incident wavepacket in the given partial wave.

To calculate this probability in the EFT, we work in the frame where the BH center of mass is at rest.   The probability is given by 
\beq
p(1\rightarrow 0) = \sum_X |{\cal A}(1+M\rightarrow 0 + X)|^2,
\eeq
where, to linear order in the interaction Eq.~(\ref{eq:sint}), the relevant matrix element is 
\beq
i{\cal A}(1+M\rightarrow 0 +X) \approx -i\int ds \langle X|Q_E^{ab}(s)|M\rangle e^i{}_a(s) e^j{}_b(s)\langle 0|E_{ij}(x^0(s),0)|\lambda\rangle + \mbox{magnetic},
\eeq
where the graviton state $\mid \lambda \rangle$ will we defined below.

Since the black hole is spinning, the change of frame $e^i{}_a(s)$ which carries co-rotating to static observers is non-trivial.   Taking the spin axis along to be along the $x^3$-axis, and denoting the angular velocity of rotation by $\Omega$, we have that $e^{\mu=0}_a=\delta^0{}_a$, and for $a=1,2,3$:
\beq
e^i{}_a(s) = \left(\begin{array}{ccc}
\cos\Omega x^0 & -\sin\Omega x^0 & 0\\
\sin\Omega x^0& \cos\Omega x^0 & 0\\
0& 0 & 1,
\end{array}\right)
\eeq
where for the black hole at rest at the origin, with momentum $p^\mu = M \delta^\mu{}_0$, Eq.~(\ref{eq:pvsx}) implies that 
\beq 
\label{mstar}
x^0(s)=M_\star s
\eeq  where $M_\star$ is an unknown constant that will drop out of the final result.

The state $|\lambda\rangle$ of the initial graviton is a linear superposition of helicity partial waves
\beq
|\lambda\rangle = \int_0^\infty {dk\over 2\pi} \psi_\lambda(k) |k,\ell,m,h\rangle,
\eeq 
which are normalized as 
\beq
\langle k,\ell,m,h|k',\ell',m',h'\rangle = 2\pi\delta(k-k') \delta_{\ell\ell'}\delta_{mm'}\delta_{hh'}.
\eeq
Then $\langle\lambda|\lambda\rangle =1$ implies that the wavefunction is normalized according to
\beq
\int_0^\infty  {dk\over 2\pi} |\psi_\lambda(k)|^2 =1.
\eeq

For graviton plane wave states, we have the matrix elements
\beq
\langle 0|B_{ij}(x^0,0)|k,h=\pm 2\rangle=\pm i\langle 0|E_{ij}(x^0,0)|k,h\rangle = \pm i{{\vec k}^2\over 2 m_{Pl}} \epsilon_{h;ij}(k) e^{-i |{\vec k}| x^0},
\eeq
where the polarization tensors of definite helicity can be expressed in terms of the $SO(3)$ Wigner $D$-matrix that takes the $z$-axis to the direction $(\theta,\phi)$ of the momentum vector ${\vec k}$,
\beq
\epsilon_{h=\pm 2,ij}(k) = \sum_{m=-2}^2 \langle i,j|\ell=2,m\rangle D^{\ell=2}_{m,h}(\theta,\phi,0)
\eeq
and $\langle i,j|\ell=2,m\rangle$  is the change of basis matrix from Cartesian to spherical rank $\ell=2$ traceless symmetric tensors\footnote{The Cartesian states are normalized as $\langle i,j|r,s\rangle={1\over 2} \left[\delta_{ir} \delta_{js} + \delta_{is}\delta_{jr} - {2\over 3} \delta_{ij}\delta_{rs}\right]$.}.   Given the relation between plane waves $|\vec k,h\rangle$ and spherical helicity eigenstates,
\beq
\langle \omega,\ell,m,h|\vec k,h'\rangle = (2\pi)^2 \sqrt{2\ell+1\over 2\pi \omega} \delta(\omega-|{\vec k}|) \delta_{hh'}  D^{\ell}_{m,h}(\theta,\phi,0),
\eeq
we can write the relevant matrix elements as
\bea
\nn
\langle 0|E_{ij}(x^0,0)|\lambda\rangle &=&\int {d^3 {\vec k}\over (2\pi)^3 2 |{\vec k}|} 2\pi\sqrt{2\ell +1\over 2\pi |{\vec k}|} \psi_\lambda(|{\vec k}|) \left[D^\ell_{m,h}(\theta,\phi,0)\right]^* \langle 0|E_{ij}(x^0,0)|k,h\rangle\\
\nn
&=&  {\sqrt{2\ell+1} \over 4 m_{Pl}}\psi_\lambda(x^0) \sum_{m'=-2}^2 \langle i,j|\ell=2,m'\rangle \int d\Omega  \left[D^\ell_{m,h}(\theta,\phi,0)\right]^* D^{\ell=2}_{m'h}(\theta,\phi,0),\\
\eea
where we have introduced the time-domain wavefunction
\beq
\psi_\lambda(x^0) = \int_0^\infty {k^{5/2} dk\over (2\pi)^{5/2}} e^{-i k x^0} \psi_\lambda(k).
\eeq
Using the orthogonality relation for the $SO(3)$ rotation matrices,
\beq
\int d\Omega \left[D^\ell_{m,h}(\theta,\phi,0)\right]^* D^{\ell=2}_{m'h}(\theta,\phi,0) = {4\pi\over 2\ell +1} \delta_{\ell,2}\delta_{mm'},
\eeq
we end up with
\beq
\langle 0|E_{ij}(x^0,0)|\lambda\rangle = {\pi\over \sqrt{5} m_{Pl}} \delta_{\ell,2}  \langle i,j|\ell=2,m\rangle \psi_\lambda(x^0),
\eeq
and $\langle 0|B_{ij}(x^0,0)|\lambda\rangle =\pm i\langle 0|E_{ij}(x^0,0)|\lambda\rangle$ as the wavepacket matrix elements with helicity $h=\pm 2$.

Since $e^i{}_a$ is a rotation matrix about the $z$-axis, it follow that
\beq
e^i{}_a e^j{}_b\langle i,j|\ell=2,m\rangle = \langle a,b|{\hat U}(R_z^{-1}(\Omega x^0)|\ell=2,m\rangle = e^{im\Omega x^0} \langle a,b|\ell=2,m\rangle,
\eeq
where ${\hat U}(R)$ is the unitary operator that represents the rotation $R$ acting on the $|\ell,m\rangle$ states.   Thus we find
\beq
\label{eq:prob}
p(1\rightarrow 0) = {\pi^2\over 5 m_{Pl}^2} \left|\int ds \langle X|Q^{ab}(s)|M\rangle \langle a,b|\ell=2,m\rangle \psi_\lambda(x^0) e^{i m\Omega x^0}\right|^2
\eeq
for the absorption probability in the EFT.

Squaring the matrix elements in Eq.~(\ref{eq:prob}) and inserting a complete set of states, 
\beq
\sum_X |X\rangle\langle X| = {\bf 1},
\eeq
relates $p(1\rightarrow 0)$ to the Wightman correlator  $\langle Q^{ab}(s) Q^{cd}(s')\rangle$ evaluated in the initial state $|M,S\rangle$ of the BH.

We would now like to write this correlator in terms of  a set of form factors which are arbitrary functions
of $\chi$. Notice that $\chi$ scales with inverse powers of $1/G_N$ and thus must be matched non-perturbatively.   The form factors are enumerated by the possible tensor structures which can now depend upon the direction of the spin, the magnitude of which is absorbed into the form factors.  It is useful to expand this correlator into a basis of tensors that are invariant under rotations about the spin axis.   Viewing the correlator as a linear map on the 5D space of traceless symmetric rank-$\ell=2$ tensors (transverse to the BH momentum $p^a$), a basis of tensors consists of the various powers of the generator $J_3$ of rotations.   Because we are in the $\ell=2$ representation of $SO(3)$, only the powers $J_3^k$ for $k=0,\ldots,4$ are independent.  For instance, $J_3^5 =  5 J_3^3 - 4 J_3$ and so on.   Thus, our tensor basis consists of the identity tensor on the $\ell=2$ space.
\beq
\langle a,b|c,d\rangle = {1\over 2} \left[\langle a|c\rangle \langle b|d\rangle +\langle a|d\rangle \langle b|c\rangle -{2\over 3} \langle a|b\rangle \langle c|d\rangle\right],
\eeq
with $\langle a|b\rangle = \delta^a{}_b - p^a p_b/p^2$  the $\ell=1$ identity matrix, together with the independent powers of the angular momentum $J_3$ in the $\ell=2$ representation.   In particular, the rotation generator in the Cartesian basis is
%
\beq
\label{eq:J3}
\langle a,b|J_3|c,d\rangle = {1\over 2} \left[\langle a|c\rangle \langle b|J_3|d\rangle +  \langle a|d\rangle\langle b|J_3|c\rangle+  \langle b|c\rangle\langle a|J_3|d\rangle+ \langle b|d\rangle\langle a|J_3|c\rangle\right]
%
%
%
\eeq
where in turn the angular momentum generator in the $\ell=1$ space is $\langle a|J_3|b\rangle= is^c\epsilon_c{}^a{}_b = {i\over M} p^c s^d\epsilon_{cd}{}^a{}_b$ (we denote the spin direction by the unit spacelike vector $s^a=\delta^a{}_3$).   The tensor $\langle a|J_3|b\rangle$ has eigenvalues $m=\pm 1,0$ corresponding to the eigenvectors $v^a_\pm = \mp{1\over \sqrt 2}\left(\delta^a{}_1 \pm i \delta^a{}_2\right)$ and $v^a_0 =s^a$ so it is normalized according to the usual conventions used in quantum mechanics.   Higher powers, of the form $\langle a,b|J_3^k|c,d\rangle$. can be obtained from Eq.~(\ref{eq:J3}) by successive tensor contraction, e.g.
\beq
\langle a,b|J_3^2|c,d\rangle = \sum_{e,f} \langle a,b|J_3|e,f\rangle\langle e,f|J_3|c,d\rangle,
\eeq
etc.   We have defined these invariant tensors such that $\langle a,b|J_3 |c,d\rangle$ is pure imaginary and Hermitian, and therefore our tensor basis satisfies the relation
\beq
\label{eq:transpose}
\langle a,b|J_3^j|c,d\rangle = (-1)^j \langle c,d|J_3^j|a,b\rangle.
\eeq
In this basis, the correlator then takes the form
\beq
\label{eq:qqexp}
\langle Q_E^{ab}(s) Q_E^{cd}(s') \rangle=M_\star^2 \sum_{j=0}^4 A^+_{E,j}(s-s') \langle a,b|J_3^j|c,d\rangle,
\eeq
where the functions $A^+_{E,j}(s-s')$ can depend on the magnitude of the particle spin as well as its mass.   We will adopt an identical decomposition for the magnetic correlator $\langle Q_B^{ab}(s) Q_B^{cd}(s') \rangle$.

In the point particle limit where our EFT is valid,  the form factors $A^+_k(s-s')$ are analytic in $\omega$, i.e. can be represented as series of derivatives acting on the delta function $\delta(s-s')$ given the lack of long time tails.  
Note that Hermiticity of the operators $Q^{ab}_{E/B}(s)$ implies that the frequency space Wightman function
\beq
W_{E/B}^{ab,cd}(\omega) = M_* \int ds e^{i\omega M_* s} \langle Q_{E/B}^{ab}(s) Q_{E/B}^{cd}(0) \rangle
\eeq
obeys the reality condition 
\beq
\label{eq:real}
[W_{E/B}^{ab,cd}(\omega)]^*= W_{E/B}^{cd,ab}(\omega)
\eeq
on the real $\omega$-axis.  Given the properties of our tensor basis, this implies that the frequency-dependent form factors $A^+_k(\omega) = M_* \int ds e^{i\omega M_* s} A^+_k(s)$ obey $[A^+_k(\omega)]^*=[A^+_k(\omega)]$ on the real axis.

Inserting the form Eq.~(\ref{eq:qqexp}) into $p(1\rightarrow 0)$, and using the fact that $J_3|\ell,m\rangle = m|\ell,m\rangle$, we obtain that
\beq
\label{eq:eftprob}
p(1\rightarrow 0) = {4\over 5} G_N {\omega^5}\sum_{j=0}^4 m^j \left(A^+_{E,j}(\omega-m\Omega)+A^+_{B,j}(\omega-m\Omega)\right).
\eeq
 The dependence on the shifted frequency $\omega-m\Omega$ reflects the transformation from the static frame to the rotating frame of the BH where the correlators are defined.   We can read off $A^+_k(\omega)$ by comparing powers of $m$ in the  result given in \cite{Starobinsky,Page} 
\beq
\label{eq:page}
p(1\rightarrow 0) \approx {16\over 225\pi} A_H (G_N M)^4\omega^5 \left[1+(m^2-1)\chi^2\right]\left[1+ {1\over 4}(m^2-4)\chi^2\right] \theta(\omega-m\Omega_H) \left(\omega-m\Omega_H\right),
\eeq
with $A_H=4\pi (r_+^2 +a^2)=8\pi (G_N M)^2\left[1+\sqrt{1-\chi^2}\right]$ the area of the horizon, $\chi = a/G_N M=J/G_N M^2$ the dimensionless rotation parameter of the Kerr black hole, and $\Omega_H = 4\pi a/A_H$ the angular velocity of the horizon.   This result is valid to all orders in the rotation parameter $\chi$, but holds to leading order in $G_N M \omega \ll 1$. The factor of $\omega-m\Omega_H,$ ensures that his result is valid in both the slow and rapidly rotating cases.    We have inserted  a step function into eq.~(\ref{eq:page}) to enforce the condition $\omega-m\Omega_H>0$ so that the single particle absorption probability is positive.  
 Naively, this seems to imply that we can not trust our results in the super-radiant regime  $\omega \ll \Omega_H$. However we can match in this regime for $m\Omega_H<0$, which can then be continued for all  $m$.

Comparison of $p(1\rightarrow 0)$ with Eq.~(\ref{eq:eftprob}) suggests that we should identify the angular velocity in the EFT with the horizon angular velocity, 
\beq
\Omega=\Omega_H= \frac{4 \pi a}{A_H},
\eeq
which, together with Eq.~(\ref{mstar}) fixes the relation between the angular velocity $\Omega_{ab}$ and spin $S^{ab}$ for a Kerr black hole,
\beq
e^{-1} \Omega^{ab} = g_{\mu\nu} e^a_\mu {D\over Ds} e^\nu{}_b = {4\pi\over A_H} {M_*\over M} S^{ab}.
\eeq
The non-vanishing frequency space response functions are then
\bea
\label{eq:a0}
A^+_{0,E}(\omega) = A^+_{0,B}(\omega)&=& {2 A_H\over 45\pi G_N} (G_N M)^4(1-\chi^2)^2 \theta(\omega)\omega,\\
\label{eq:a2}
A^+_{2,E}(\omega) = A^+_{2,B}(\omega)&=& {A_H\over 18\pi G_N} (G_N M)^4\chi^2(1-\chi^2) \theta(\omega) \omega,\\
\label{eq:a4}
A^+_{4,E}(\omega) = A^+_{4,B}(\omega)&=& { A_H\over 90\pi G_N} (G_N M)^4\chi^4 \theta(\omega) \omega.
\eea
In obtaining this result, we have assumed equality of the electric and magnetic response.  We will check this assumption below by comparing to known results obtained via different methods.

The step function $\theta(\omega)$ reflects that matching was performed under the assumption the graviton is quantized around the Boulware vacuum~\cite{boulware}, corresponding to no (Hawking) particle emission for $\omega - m \Omega_H>0$.   By contrast, matching in the Unruh state~\cite{unruh}, where the BH can emit Hawking radiation,  would lead to Wightman response functions $A_+(\omega)$ that are non-vanishing even at $\omega <0$.   See~\cite{Houses} for a more detailed discussion of matching in the Unruh state.   It is straightforward to check that in the Boulware state, the single particle emission probability $p(0\rightarrow 1)$ is given in the EFT by a formula like Eq.~(\ref{eq:eftprob}) that involves the Wightman correlators $A_+(m\Omega-\omega)$, leading to the prediction of a non-zero emission probability in the for superradiant modes with $\omega-m\Omega_H<0$.

\beq
p(0\rightarrow 1) \approx {16\over 225\pi} A_H (G_N M)^4\omega^5 \left[1+(m^2-1)\chi^2\right]\left[1+ {1\over 4}(m^2-4)\chi^2\right] \theta(m\Omega_H-\omega) \left(m\Omega_H-\omega\right),
\eeq
see~\cite{penco2} for more a detailed discussion of the worldline EFT in the regime of superradiant emission.


\subsection{The causal response function}
\label{sec:ret}

In the classical processes that we consider in this paper, the relevant correlator is the retarded Green's function
\beq
G_R^{ab,cd}(s-s') = -i\theta(s-s') \langle \left[Q^{ab}(s),Q^{cd}(s')\right]\rangle
\eeq
rather than the Wightman functions obtained in the previous section.   Because this is a real quantity, the frequency space causal response $G^{ab,cd}_R(\omega) = M_* \int ds e^{i\omega M_* s}G^{ab,cd}_R(s)$ satisfies the reality condition 
\beq
\left[G^{ab,cd}_R(-\omega)\right]^*=G^{ab,cd}_R(\omega),
\eeq
for real frequencies.   Thus  $\mbox{Re}  G^{ab,cd}_R(\omega)$ is an even function on the real $\omega$-axis while $\mbox{Im}  G^{ab,cd}_R(\omega)$ is an odd function.   The retarded Green's function is related to the two-point Wightman correlators by a dispersion relation of the form
\beq
\label{eq:gret}
G^{ab,cd}_R(\omega) = M_* \int ds e^{i\omega M_* s}G^{ab,cd}_R(s) =\int_{-\infty}^\infty {d\omega'\over 2\pi} {W^{ab,cd}(\omega') - W^{cd,ab}(-\omega')\over \omega-\omega'+i\epsilon},
\eeq
which, as a consequence, defines a function that is analytic for $\mbox{Im}\omega\geq 0$ but singular on the lower-half complex-$\omega$ plane.  Expanding out the dispersion relation in Eq.~(\ref{eq:gret}) into its real and imaginary parts, we find that in terms of the Wightman functions
\bea
\nn
\mbox{Re} G^{ab,cd}_R(\omega) = {1\over 2} \mbox{Im} \left[W^{ab,cd}(\omega) - W^{cd,ab}(-\omega)\right] +\mbox{Pr}\int_0^\infty {\omega' d\omega'\over \pi} { \mbox{Re} \left[W^{ab,cd}(\omega') - W^{cd,ab}(-\omega')\right]\over \omega^2-\omega'^2}\\,
\label{eq:ReG}
\eea
and
\bea
\nn
\mbox{Im} G^{ab,cd}_R(\omega) = -{1\over 2} \mbox{Re} \left[W^{ab,cd}(\omega) - W^{cd,ab}(-\omega)\right] +\omega\cdot \mbox{Pr}\int_0^\infty { d\omega'\over \pi} { \mbox{Im} \left[W^{ab,cd}(\omega') - W^{cd,ab}(-\omega')\right]\over \omega^2-\omega'^2}.\\
\label{eq:ImG}
\eea
This result follows from Eq.~(\ref{eq:real}), which implies the exchange properties under the transformations $\omega\rightarrow-\omega$ or $ab\leftrightarrow cd$ listed in Table~\ref{tab:ex}.

\begin{table}[t]
\centering
\begin{tabular}{c|c|c}
 & $\omega\leftrightarrow -\omega$ &$ab\leftrightarrow cd$\\
 \hline
$\mbox{Re} \left[W^{ab,cd}(\omega) - W^{cd,ab}(-\omega)\right]$  & odd & even\\
& &\\
 $\mbox{Im} \left[W^{ab,cd}(\omega) - W^{cd,ab}(-\omega)\right]$  & even & odd\\
 \end{tabular}
\centering
 \caption{Behavior of real and imaginary parts under the substitution $\omega\rightarrow -\omega$ or index exchange $ab\leftrightarrow cd$.   Even/odd means that the function changes/does not change sign under the given transformation.}
\label{tab:ex}
\end{table}

Note that in addition to the contribution of the worldline multipole operators, the physical response (as determined, for example, through measurements of the gravitational field at large distances) can also receive contributions from  terms in the worldline action that are polynomial in $E_{ab}$, $B_{ab}$ and/or their derivatives with respect to the parameter $s$.  We will henceforth refer to these
terms as ``local'', to make the distinction  from terms in the action involving the internal degrees of freedom $X$.

  Focusing on purely electric couplings, such local terms modify the low-frequency response by an analytic function $L^{ab,cd}(\omega)$ whose real part is \emph{even} under $\omega\leftrightarrow -\omega$ or index interchange $ab\leftrightarrow cd$.   Then using  (\ref{eq:transpose}) we see that  the local contribution to the real response can only involve the tensor structures $\langle a,b|J_3^j|c,d\rangle$ with $j=0,2,4$.  In particular, the  the static Love numbers of the black hole, which are  identified with the local response at $\omega=0$ (both from $L^{ab,cd}(\omega)$ and from Eq.~(\ref{eq:gret})) cannot involve tensor structure that are linear or cubic in the spin. Alternatively, 
time reversal invariance implies that terms odd in spin vanish in the static limit. 

The contribution from terms in the action also modifies the imaginary part by terms that are odd under either $\omega\rightarrow -\omega$ or $ab\leftrightarrow cd$ exchange\footnote{An example of such term is Eq.~(\ref{eq:localedot}) below.}.   Despite possibly having a non-vanishing  imaginary part, the local response $L^{ab,cd}(\omega)$ does not contribute to dissipation, as will discussed below,    and thus can not be matched using $p(1\rightarrow 0)$ but instead  must be fixed by matching to other observables in the full theory, for instance elastic scattering of low-frequency gravitons off the black hole.

 Because our matching procedure only fixes the Wightman function at low frequency, it does not completely determine the form of the retarded response function.  In particular, matching to low-frequency absorption cannot yield information about the terms in Eq.~(\ref{eq:ReG}) and Eq.~(\ref{eq:ImG}) that involve principal part integrals over high arbitrarily high frequency scales, where the EFT description necessarily breaks down.   However, from Eq.~(\ref{eq:ReG}) and  Table~\ref{tab:ex}, we see that the principal part integral contribution to $\mbox{Re} G_R^{ab,cd}(\omega)$ is analytic at $\omega=0$ (assuming the integral in Eq.~(\ref{eq:ReG}) converges), and even under either $\omega\leftrightarrow -\omega$ or $ab\leftrightarrow cd$ exchange.   Similarly $\mbox{Im} G_R^{ab,cd}(\omega)$ is odd if we replace  $\omega\leftrightarrow -\omega$ or $ab\leftrightarrow cd$.  Consequently, the principal part contribution to $G_R^{ab,cd}(\omega)$ is physically indistinguishable (i.e. of the same form), from the local response $L^{ab,cd}(\omega)$ arising from adding local counterterms to the point particle action.   
 
 On the other hand, the calculable part of Eqs.~(\ref{eq:ReG}),  Eqs.~(\ref{eq:ImG}) gives rise to a genuinely non-local contribution to the retarded Green's function, of the form
  \beq
 G^{ab,cd}_{R,non-local}(\omega) =  -{i\over 2}  \left[W^{ab,cd}(\omega) - W^{cd,ab}(-\omega)\right].
 \eeq
This object does not have the correct $ab\leftrightarrow cd$ index exchange properties to arise from curvature couplings in the point particle action, and cannot be absorbed into a local counterterm.   It is in particular this function $G^{ab,cd}_{R,non-local}(\omega)$ that gives rise to dissipative effects in the EFT description of the black hole.

Ignoring the local contribution to the causal response, we obtain from Eqs.~(\ref{eq:a0})-(\ref{eq:a4}) the result
\beq
\label{eq:gres}
G_{R,E}^{ab}{}_{cd}(\omega) = {M^2 A_H\over 45\pi G_N} (G_N M)^4 \left(-i\omega\right)\cdot \langle a,b|  (1-\chi^2)^2 + {5\over 4} \chi^2 (1-\chi^2) J_3^2+ {1\over 4}\chi^4 J_3^4|c,d\rangle,
\eeq   
with an identical expression for the magnetic Green's function $G_{R,B}^{ab}{}_{cd}(\omega)$.   This result is equivalent to the statement that, up to local terms, the quadrupole moment induced\footnote{Because of spin, the Kerr black hole has an infinite series of permanent multipole moments~\cite{hansen}, which in the point particle limit are equivalent to local spin-dependent worldline interactions that linearly in the curvature tensor.   Here, by induced moment, we mean the shift in the value of the permanent moments that are generated when a background field $R_{\mu\nu\rho\sigma}\neq 0$ is turned on.} by an external electric field is
\bea
\label{Qres}
\nn
\langle Q^{ab}_E(s)\rangle  &=& \int ds' G_{R,E}^{ab}{}_{cd}(s-s') E_{cd}(s')\\
\nn
&=& { A_H\over 45\pi G_N} (G_N M)^4  \langle a,b|  (1-\chi^2)^2 + {5\over 4} \chi^2 (1-\chi^2) J_3^2+ {1\over 4}\chi^4 J_3^4|c,d\rangle {d\over ds} E_{cd}(x(s)),\\
\label{eq:Qind}
\eea
where the derivative here is in the co-rotating frame, see Eq.~(\ref{eq:edot}).
 An identical formula relates the induced magnetic moment $\langle Q^{ab}_E(s)\rangle$ to the co-rotating components of magnetic curvature $B_{ab}(x(s))$ along the point particle worldline.
 
Taking the limit where the rotation of the black hole is larger than the intrinsic time dependence of the curvature, we may approximate
\beq
{d\over ds} E_{ab}\approx -\Omega_a{}^c E_{cb} - \Omega_b{}^c E_{ac}= {i} M_*\Omega_H \langle a,b|J_3|c,d\rangle E_{cd},
\eeq
in which case the induced moment is of the form 
\beq
\label{eq:drop}
M_*^{-1} \langle Q^{ab}_E(s) \rangle \approx {4 i (G_N M)^5\over 45 G_N}\chi\langle a,b|(1-2\chi^2) J_3 + {5\over 4} \chi^2 J_3^3 |c,d\rangle {E}_{cd}.
\eeq
Despite appearances, Eq.~(\ref{eq:drop}) does not imply the existence of a non-vanishing static Love tensor for the Kerr black hole, since this relation cannot arise from local terms in the point particle action.   In particular, a term such as $\int \chi ds E_{ab} \langle a,b|J^j_3|c,d\rangle E_{cd}$ for $j=1,3$ vanishes identically due to the antisymmetry under $ab\leftrightarrow cd$ of the tensor structures.    We have verified, however, that Eq.~(\ref{eq:drop}) is consistent, for $\chi\ll 1$, with the results of~\cite{LeTiec:2020spy} which obtained the quadrupolar response of a slowly spinning Kerr BH, at linear order in $\chi$.

On the other hand, it is in principle possible that Kerr black holes have non-zero static Love numbers, but by symmetry those would have to correspond to local worldline counterterms which in our basis take the form
\bea
\label{eq:cts}
S_{pp}  \supset G_N^4 M^6\int ds f_j(\chi^2) \chi^j E_{ab} \langle a,b|J_3^j|c,d\rangle E_{cd},
\eea
with $j=0,2,4$, as well as their magnetic counterparts.   Here, we have defined $\chi ={\sqrt{-S^\mu S_\mu}\over G_N p^2}\leq 1$, and $f_{0,2,4}(\chi^2)$ are functions analytic at $\chi^2=0$.    The overall scaling $G_N^4 M^6$ is the characteristic magnitude of the static tidal response of a compact object.   It is well known that the spin-independent term in Eq.~(\ref{eq:cts}) has vanishing Wilson coefficient, i.e $f_0(\chi^2=0)=0,$~\cite{Binnington:2009bb,Damour:2009va,Kol:2011vg}.   Recently, ref.~\cite{Chia:2020yla} has extended this calculation to arbitrary orders in spin (previous partial results can be found in~\cite{Pani:2015hfa}) and found, remarkably, that the all local contributions to the static response function of the Kerr BH are in fact vanishing as well.

In addition to the local contributions to the static response, there are also terms in the point particle action which modify $G_R^{ab,cd}(\omega)$ away from $\omega=0$.   The leading such terms at low frequency are of the form
\beq
\label{eq:localedot}
S_{pp} \supset G_N^5 M^6\int ds \chi^j E_{ab}\langle a,b| [iJ_3]^j|c,d\rangle {\dot E}_{cd},
\eeq
with $j=1,3$.   These local interactions not forbidden by symmetries (it is even under both parity and time reversal), and yield contributions to $\langle Q_E^{ab}\rangle$ of comparable magnitude to those in Eq.~(\ref{eq:Qind}).  However, unlike the terms in Eq.~(\ref{eq:Qind}), the curvature couplings in Eq.~(\ref{eq:localedot}) cannot give rise to dissipative effects, despite the fact that they contribute to $\mbox{Im} G_{R,E}^{ab,cd}(\omega)$.     The recent analysis of ref.~\cite{Chia:2020yla}  indicates that, for the Kerr BH, terms such as those in Eq.~(\ref{eq:localedot}) are also vanishing.   Assuming the validity of the results in~\cite{Chia:2020yla}, it then follows that Eq.~(\ref{eq:gres}) completely characterizes the black hole response function at linear order in time derivatives but to all orders in spin.

\subsection{Check:  Dissipative dynamics of a black hole in a tidal environment}
\label{sec:poi}
Assuming that all the local contributions to BH response are indeed zero~\cite{Chia:2020yla}, the complete equations of motion for a spinning black hole moving in a background gravitational field with curvature scale ${\cal R}\gg G_N M$ can be obtained straightforwardly by inserting the induced moments $\langle Q_{E,B}\rangle$ from Eq.~(\ref{eq:Qind}) and its magnetic analog into Eqs.~(\ref{eq:dpdt}),~(\ref{eq:dsdt}).   Because the resulting expressions are messy and not particularly illuminating, we will report instead on the implications of these equations for the rate of change of mass and spin that arise as a consequence of tidal interactions, given in Eqs.~(\ref{eq:dmdt}),~(\ref{eq:ds2dt}).    We consider separately the cases of a rapidly spinning BH,  ${\cal R}^{-1}\ll \Omega_H$ and $\chi\sim{\cal O}(1)$, as well as the opposite slow-spin limit ${\cal R}^{-1}\gg\Omega_H$, which necessarily requires that $|\chi|\ll 1$,

Using the relation in Eq.~(\ref{mstar}) between our parameter $s$ and the proper time $\tau$ along the wordlline of the rotating BH, we find, in the large spin case
 \bea
 \nn
 {d\over d\tau} M &\approx& {8 (G_N M)^5\over 45 G_N}\chi \epsilon^{\mu\nu}{}_\lambda s^\lambda\left[(1+3\chi^2) E_{\mu\rho}{\dot E}^\rho{}_\nu  + {15\over 4}\chi^2 E_{\mu\rho} s^\rho {\dot E}_{\nu\sigma} s^\sigma\right]+\mbox{magnetic}\\
& & + {\cal O}(G_N M/{\cal R}).
\label{eq:fast}
\eea
This result, which is valid to all orders in spin, agrees with result found in ref.~\cite{poisson2,poisson3}.  To linear order in $\chi$ it also agrees with results obtained in~\cite{Porto}.     In the opposite, $\chi\rightarrow 0$ limit, we find instead
\bea
{d\over d\tau} M &\approx& {16 \over 45 G_N} (G_N M)^6\left[{\dot E}_{\rho\sigma} {\dot E}^{\rho\sigma} +{\dot B}_{\rho\sigma} {\dot B}^{\rho\sigma} \right] +  {\cal O}(\chi)
\label{eq:slo}
\eea  
which receives corrections at linear order in $\chi\ll 1$ from radiative tail contributions to the EFT matching and to the Schwinger-Keldysh action.   This is also in agreement with~\cite{death,Poisson:2004cw}.

  For the torque induced on the black hole by the tidal background, we find from Eq.~(\ref{eq:dsdt}),~(\ref{eq:Qind}) 
  \bea
  \label{eq:torquehi}
  \nn
  {d\over d\tau} S &\approx& -{2\over 45 G_N} (G_N M)^5\chi\left[8 (1+3\chi^2) E_{\rho\sigma} E^{\rho\sigma} + 3 (4+17\chi^2) E_{\lambda\rho} E^{\lambda}{}_\sigma s^\rho s^\sigma + 15\chi^2 (E_{\rho\sigma} s^\rho s^\sigma)^2\right]\\
 \nn\\
 & & {}\hspace{0.5cm}+\mbox{magnetic}.  
  \eea
  in the limit $\Omega_H\gg {\cal R}^{-1}$.   In the opposite, $\chi\rightarrow 0$, Eq.~(\ref{eq:Qind}) is dominated by the time variation of $E_{\mu\nu}$ and the torque is instead
  \bea
  \label{eq:torquelo}
  {d\over d\tau} S &\approx& - {8\over 45 G_N} (G_N M)^6 \epsilon^{\mu\nu}{}_\lambda s^\lambda \left[E_{\mu\rho}{\dot E}^\rho{}_\nu+B_{\mu\rho}{\dot B}^\rho{}_\nu\right]
  \eea
  Both Eqs.~(\ref{eq:torquehi}),~(\ref{eq:torquelo}) are in agreement with results obtained previously in~\cite{death,Poisson:2004cw}.  To go to next order in $G_N M/{\cal R}\ll 1$ would require the inclusion in both the EFT matching and Schwinger-Keldysh action of infrared divergent tail terms corresponding to graviton scattering off the BHs own gravitational field.   Ref.~\cite{poisson3} has reported a result for these next-to leading order corrections, although a discrepancy with their earlier results~\cite{poisson2} obtained in a probe limit remains unsettled in the literature.


As another check of our results, note that from Eqs.~(\ref{eq:dpdt}),~(\ref{eq:dsdt}), we also find that in terms of the curvatures $E_{ab}, B_{ab}$ in the rotating frame
\beq
{d\over d\tau} M-\Omega_H {d\over d\tau} S = \langle Q^E_{ab} \rangle {D\over D\tau} E^{ab} +  \langle Q^B_{ab}\rangle {D\over D\tau} B^{ab},
\eeq
or by Eq.~(\ref{eq:Qind}),
\bea
\nn
{d\over d\tau} M-\Omega_H {d\over d\tau} S &=& { A_H (G_N M)^4 \over 45\pi G_N}   {d\over d\tau} E_{ab} \langle a,b|  (1-\chi^2)^2 + {5\over 4} \chi^2 (1-\chi^2) J_3^2+ {1\over 4}\chi^4 J_3^4|c,d\rangle {d\over d\tau} E_{cd}\\
\nn\\
& & {} + \mbox{magnetic.}
\eea 
Because the even powers of the tensor $\langle a,b|J_3|c,d\rangle$ are positive definite, this quantity is manifestly positive in the physical region $\chi^2 \leq 1$.  Therefore the change in the BH area as a result of tidal interactions is also positive
 \bea
 {d\over d\tau} A_H = {2 A_H\over M\sqrt{1-\chi^2}}\left[{\dot M} - \Omega_H {\dot S}\right] \geq 0,
 \eea
 as required on general grounds~\cite{hawking}.

 \section{Post-Newtonian equations of motion for binary dynamics}
 \label{sec:PN}
 
 \begin{figure}[t]
    \centering
    \includegraphics[scale=0.22]{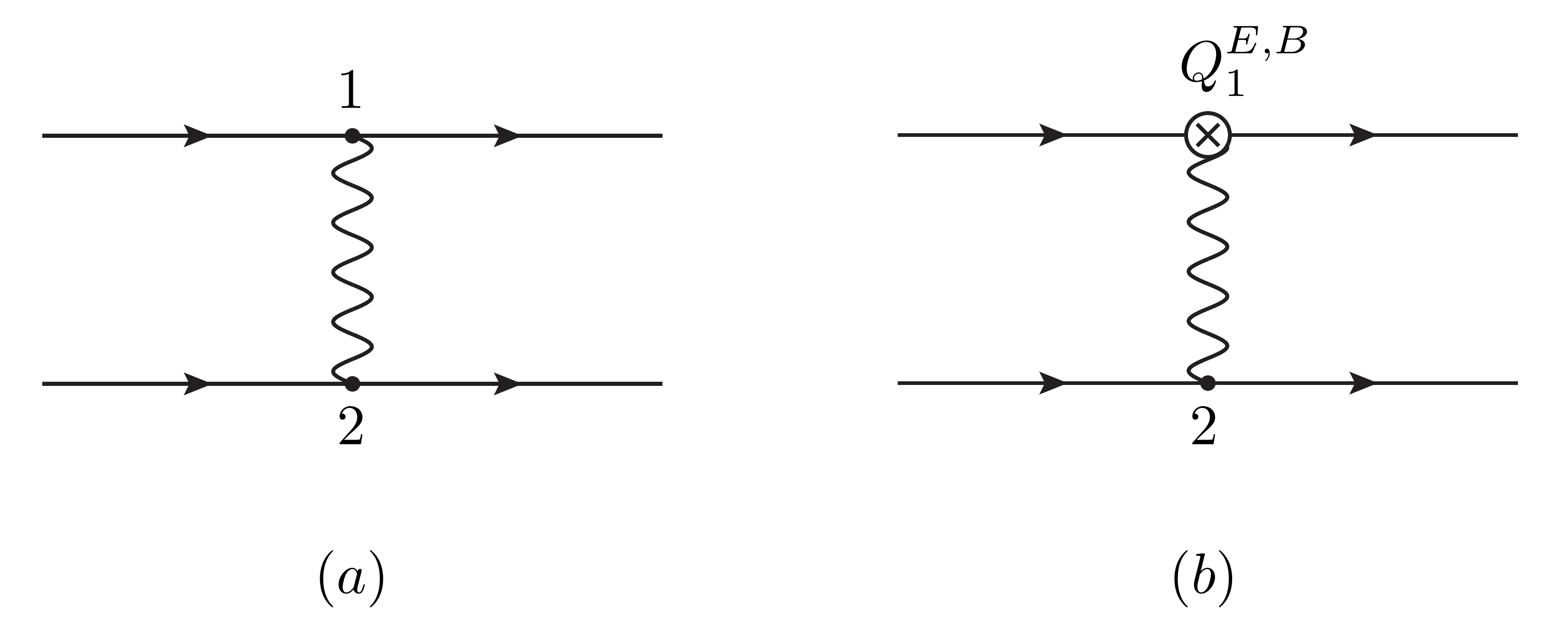}
\caption{Potential exchange diagrams that contribute to the two-particle action $S_{int}$.   In (a) the particles interact through the minimal gravitational interaction.    Figure (b) is the term in $S_{int}$ generated by the quadrupole couplings of particle 1 (a similar diagram with $1\leftrightarrow 2$ has been omitted).}
\label{fig}
\end{figure}

 The same worldline effective action formalism can also be applied to dissipation in dynamically generated spacetimes, i.e sourced by the particles themselves, rather than the fixed background field case discussed above.   In order to do so, we have to include in the Schwinger-Keldysh functional an integral over the fluctuations of the gravitational field itself\footnote{The role of the in-in formalism to describe radiation reaction forces was first discussed in~\cite{galley}.}.

 As an example we will consider a binary system of black holes in the non-relativistic regime, with $v^2\sim G_N Mr/r \ll 1$.   For illustration, we will focus on the regime of rapidly spinning black holes, with $\Omega_H\gg v/r$.   The rotation parameters will be assumed to scale as $\chi\sim{\cal O}(1)$.    Integrating out the potential graviton exchange between the black holes, Fig.~\ref{fig}(b), the two-particle interaction term reduces to \cite{GnR3}  
 \beq
 S_{int} \approx -G_N m_1 m_2 \int dt  \left[{Q^{ab}_{E,1}(t)\over m_1^2} {e_1}^i{}_a {e_1}^j{}_b +(1\leftrightarrow 2)\right] \partial_i \partial_j {1\over |{\vec x}(t)|},
 \eeq
 with ${\vec x}={\vec x}_1-{\vec x}_2$, up to terms suppressed by more power of the velocities.   Varying the in-in action, we obtain, in the linear response limit, an instantaneous non-conservative force on the black holes that is given by,
\bea
\nn
{\vec F}_1(t) &=& {\delta\over \delta {\vec x}_1(t)}\left. \Gamma[{\vec x},\tilde{\vec x};e_{1,2},{\tilde e}_{1,2}]\right|_{\vec x=\tilde{\vec x};e_{1,2}={\tilde e}_{1,2}}\\
\nn
&\approx& -G_N m_1 m_2  \left[{\langle Q^{ab}_{E,1}(t)\rangle\over m_1^2}  {e_1}^a{}_j {e_1}^b{}_k+(1\leftrightarrow 2)\right]
\nabla \partial_j \partial_k {1\over |{\vec x}(t)|} = -{\vec F}_2(t),\\
\eea
with ${\vec x}={\vec x}_1-{\vec x}_2$.   Similarly, the torque on each black hole can be obtained from the Schwinger-Keldysh action by varying with respect to the frame $e^a{}_i$,  
\bea
\nn
{d\over dt} {\vec S}_1^i = e_1^i{}_a \epsilon^{abc}  {\delta\over \delta\theta_1{}^{bc}} \left. \Gamma[{\vec x},\tilde{\vec x};e_{1,2},{\tilde e}_{1,2}]\right|_{\vec x=\tilde{\vec x};e_{1,2}={\tilde e}_{1,2}} \approx  {2 G_N m_2\over m_1} e_1^i{}_a e_1^j{}_b e_1^k{}_c \epsilon^{abd} \langle Q^{E,1}{}^c{}_d \rangle\partial_j \partial_k |{\vec x}|^{-1}.
\\
\eea
On the right hand side of this and the previous equation, the in-in expectation values in the PN limit can be obtained from Eq.~(\ref{eq:Qind}), by inserting
\beq
E_{ab}= G_N m_2 {e_1}^i{}_a  {e_1}^i{}_b \partial_i \partial_j |{\vec x}(t)|^{-1}
\eeq
into $\langle Q^{ab}_{E,1}\rangle$, and similarly for the case of $\langle Q^{ab}_{E,2}\rangle$.   This yields the result
\bea
\label{eq:for}
{\vec F}_1(t)= -{\vec F}_2(t)=-{8\over 5} {G^5_N m_1^3 m_2^2\over |{\vec x}|^7}\left[1+3\chi_1^2 - {15\over 4} \chi_1^2 \left({\vec s}_1\cdot {{\vec x}\over |{\vec x}|}\right)^2\right]  {{\vec x}\over |{\vec x}|}\times {\vec S}_1 +(1\leftrightarrow 2)
\eea
for the non-conservative force.  The torque on each particle is
\bea
\label{eq:tor}
{d\over dt} {\vec S}_1 =  -{8\over 5} {G^5_N m_1^3 m_2^2\over |{\vec x}|^6}\left[1+3\chi_1^2 - {15\over 4} \chi_1^2 \left({\vec s}_1\cdot {{\vec x}\over |{\vec x}|}\right)^2\right]  \left[{\vec S}_1 -  {{\vec S}_1\cdot{\vec x}\over {\vec x}^2} {\vec x}\right].
\eea
In Eq.~(\ref{eq:for}), ``$1\leftrightarrow 2$" has the meaning that we exchange the particle labels without changing the sign of ${\vec x}$.   The PN equations of motion to linear order in the spin for an arbitrary composite object were first calculated in \cite{Endlich:2015mke}.    Our results at $\chi\ll 1$ agree with those of ref.~\cite{Endlich:2015mke} if one uses  Eq.~(\ref{eq:a0}) with $\chi=0$ to fix their dissipation parameter.   The friction force ${\vec F}_{1,2}$ is a 5PN effect, while our result for the torque is 4PN relative to the leading order gravito-magnetic spin precession formula predicted by linearized GR.   

As a simple consequence of Eq.~(\ref{eq:for}), consider the mechanical power that is absorbed or extracted by the BH horizons
\beq
\label{eq:PNpow}
{d\over dt} E =\sum_a {\vec v}_a\cdot {\vec F}_a \approx {8\over 5} {G^5_N m_1^2 m_2\over |{\vec x}|^8} (m_1+m_2) \left[1+3\chi_1^2 - {15\over 4} \chi_1^2 \left({\vec s}_1\cdot {{\vec x}\over |{\vec x}|}\right)^2\right] {\vec S}_1\cdot {\vec L}+(1\leftrightarrow 2),
\eeq
where ${\vec L}$ is the orbital angular momentum about the center of mass. This results agrees to linear order with \cite{Porto,Endlich}.  Depending on the relative orientations between the spins and the orbit, the rate of change of energy can be positive or negative, reflecting the possibility of energy extraction from the black holes through the Penrose process.   For example, if the spins are orthogonal to the orbital plane, $dE/dt$ can be either positive or negative depending on whether the spins are aligned or anti-aligned with ${\vec L}$.   Regardless, Eq.~(\ref{eq:PNpow}) enters at order $v^5$, or 2.5PN relative to leading order quadrupole radiation from the binary and, as is well known~\cite{Tagoshi:1997jy,Poisson:2004cw}, is enhanced relative to absorption in the case of non-rotating black holes by a factor of $v^{-3}$.    A final check of these results is that the orbital angular momentum as predicted by Eq.~(\ref{eq:for}) is given by 
\beq
{d\over dt} {\vec L} = \sum_a {\vec x}_a \times {\vec F}_a \approx -{d\over dt}( {\vec S}_1 +{\vec S}_2),
\eeq
with $d{\vec S}_{1,2}/dt$ given by Eq.~(\ref{eq:tor}).   It therefore follows that the total angular momentum ${\vec J}={\vec L}+{\vec S}_1 + {\vec S}_2$ is conserved, as should be expected given that the tidal dynamics we consider here does not involve any gravitational radiation out to infinity at leading PN order.

\section{Conclusions}
\label{sec:conc}

In this work we have calculated the leading order non-conservative finite size effects in Kerr black hole dynamics within 
the worldline EFT formalism.   In contrast to earlier approaches~\cite{Porto,Endlich}, the EFT is valid for arbitrary rotation parameter $\chi$ within the physical region $\chi^2\leq 1$.   Using our framework, we have obtained results for angular momentum and energy loss in a background field which agree, upon time averaging, with those previously obtained in refs.~\cite{Poisson:2004cw,poisson2,poisson3}.   

We have also presented results for the 5PN equations of motion of near extremal black holes, as well as the 2.5PN correction to the power transferred between horizon and orbital degrees of freedom.    Due to the large size of this latter effect, the next to leading order PN non-conservative effects are phenomenologically relevant as well.    Part of the motivation for the present work has been to set up a systematic method which, combined with~\cite{GnR1}, can be used to calculate such corrections.   One potential application is to resolve the discrepancy between results obtained in~\cite{Tagoshi:1997jy} and the the test particle limit of formulas obtained in~\cite{poisson3} for general mass ratios.   Part of the discrepancy between the two results lies in certain terms proportional to $\pi^2$ which, in the worldline EFT,  arise from infrared enhanced tail-type corrections.   We hope to address these effects in future work.

{\bf Note added}:  Some of these results were first reported in the talk~\cite{talk}.   While this paper was in preparation, ref.~\cite{LeTiec:2020bos} appeared which has partial overlap with work reported here and in \cite{talk}.  In particular Eq.~(\ref{Qres}) appears as the Love tensor given in Eq.~(8.10) of \cite{LeTiec:2020bos}.

\section{Acknowledgments}

This work was partially supported by the US Department of Energy under grants DE-SC00-17660 (WG and JL) and DE- FG02-04ER41338 and FG02- 06ER41449 (IZR).

\end{document}